\documentclass[%
reprint,
amsmath,amssymb,
aps,prd,
showkeys
]{revtex4-2}

\usepackage{graphicx}
\usepackage{dcolumn}
\usepackage{bm}
\usepackage{hyperref}
\hypersetup{
    colorlinks,
    citecolor=red,
    filecolor=black,
    linkcolor=blue,
    urlcolor=black
}
\usepackage{nicematrix}
\usepackage{tikz}
\usetikzlibrary{quantikz2}
\usepackage{orcidlink}

\usepackage[arrowdel]{physics}
\newcommand{\eps}{\varepsilon}

\begin{document}

\title{Feynman's clock and hierarchy-informed sampling for quantum error mitigation}
\date{\today}

\author{Theo Saporiti \orcidlink{0009-0008-9738-3402}}
\email{theo.saporiti@cea.fr}
\affiliation{Université Paris-Saclay, CEA, List, F-91120, Palaiseau, France}

\begin{abstract}\noindent
Near-term physical implementations of quantum algorithms require efficient quantum error mitigation schemes to reduce quantum noise. In this letter we propose a new mitigation technique, by extending the applicability of our BBGKY-ISM scheme from quantum simulations of spin chains to arbitrary quantum circuits. We map executions of quantum circuits using Feynman's clock Hamiltonian to the Hamiltonian dynamics of a corresponding quantum system, whose time evolution obeys a BBGKY-like hierarchy of equations informing the BBGKY-ISM mitigation. We show that the method's classical and quantum overheads are polynomial in the circuit size and in the number of qubits. We apply our method to numerical simulations of tunable Bell state preparation circuits under state-of-the-art quantum noise, and numerically demonstrate its systematic and controllable quantum error reduction capability.
\end{abstract}


\maketitle

\section{Introduction}

It is known that some quantum algorithms, such as Grover's \cite{Grover:1996rk} or the Deutsch–Jozsa's algorithms \cite{Deutsch:1992idk}, can theoretically outperform their classical counterparts. However, the actual physical execution of quantum circuits on present quantum computers is spoiled by their noisy intermediate-scale quantum (NISQ) nature \cite{Steane:1997kb, DiVincenzo:2000tra, Ladd:2010cup, Georgescu:2013oza, Preskill:2018jim}. Pending the large-scale devices required for the application \cite{Chatterjee:2024kpt} of error correction techniques \cite{ShorEC, Steane, Calderbank, Nielsen_Chuang_2010}, efficient quantum error mitigation schemes were developed \cite{PhysRevApplied.20.064027, PhysRevA.94.052325, PhysRevA.105.032620, Temme, Li:2016vmf, Giurgica-Tiron:2020rcf, PracticalQEM, LearningBasedQEM, AIQEM, Cai:2020khs, PhysRevA.104.052607, PhysRevResearch.3.033098, PhysRevA.105.042408, Montanaro:2021amf, article10313813, Czarniketal2021, Saporiti:2025zwf, Kaikov:2025dtr, Saporiti:2026lbg}. Although fundamentally limited \cite{FundamentalLimits, PhysRevLett.131.210602, FundamentalLimits2}, they are the only quantum error reduction techniques readily deployable for near-term quantum applications \cite{naturearticle, Bultrini2023, RevModPhys.95.045005}, and they are expected to hold importance even at the beginning of the fault-tolerant era \cite{PRXQuantum.3.010345, Zimboras:2025unr, Zhang:2025gyl}.

In this letter we propose a new quantum error mitigation technique, capable of mitigating the quantum noise stemming from the execution of any quantum circuit. Our proposed method extends the applicability of the BBGKY Informed Sampling Mitigation (BBGKY-ISM) scheme, previously employed on noisy simulations of spin chain systems \cite{Saporiti:2026lbg}, to mitigate arbitrary noisy circuits. The core idea behind the technique is to map circuits to a corresponding quantum system, given by Feynman's quantum computer construction \cite{Feynman:1984bi, Costales:2023ljr, Caha:2018yhv}, and mitigate quantum simulations of its time evolution using BBGKY-ISM. The latter samples possible mitigation candidates from a physics-informed probability distribution, encoding the noiseless dynamics in the form of Bogoliubov-Born-Green-Kirkwood-Yvon (BBGKY) hierarchical equations. A remarkable feature of our proposed method is that it can be easily coupled with existing mitigation schemes, in that it is an entirely post-processing technique.

We now outline the contents of this letter. In Sec.~\ref{sec:technique} we present our proposed technique, namely we first give a brief overview of Feynman's quantum computer (Sec.~\ref{subsec:feynman}), we then build the connecting procedure from the former to BBGKY-ISM, together while showing the efficient scaling of both classical and quantum overheads (Sec.~\ref{subsec:alternativeZ}), and finally we provide a brief reminder of the BBGK-ISM inner workings (Sec.~\ref{subsec:bbgkism}). In Sec.~\ref{sec:results} we test the method on the concrete example of the tunable Bell state preparation circuit, numerically demonstrating the systematic error reduction capability of our scheme. Finally in Sec.~\ref{sec:conclusions} we conclude and provide further expansions of this work.

\section{Mitigation technique}\label{sec:technique}

Consider an arbitrary $N_\text{Q}$-qubits circuit $U$ built as a sequence of $N_\text{G}$ quantum gates $U_{g \in \qty{1,\dots,N_\text{G}}} \in \mathcal{S}$, all picked from a chosen universal quantum gate set $\mathcal{S}$. Then $U := U_{N_\text{G}} U_{N_\text{G}-1} \cdots U_1 U_0$, where $U_0 := \text{id}$ is the empty circuit, and we say that $U$ acts on quantum states of the data register $\mathcal{H}_\text{D} := \text{span}_{\mathbb{C}}\qty{\ket{0},\ket{1}}^{\otimes N_\text{Q}}$. Given the initial input state $\ket{\phi}_\text{D} \in \mathcal{H}_\text{D}$, by defining the computational states $\ket{g}_\text{D} := U_{g} U_{g-1} \cdots U_0 \ket{\phi}_\text{D}$ for any $0 \leq g \leq N_\text{G}$, the result of the computation encoded in $U$ is $\ket{N_\text{G}}_\text{D}$, and $N_\text{Q}$ individual Z-measurements can be performed on its corresponding qubits \cite{Costales:2023ljr}.

\subsection{Feynman's quantum computer}\label{subsec:feynman}
The execution of a quantum circuit in the above form can be recast into the time evolution of a quantum system whose time-independent Hamiltonian acts in an augmented Hilbert space $\mathcal{H}_\text{C} \otimes \mathcal{H}_\text{D}$ \cite{Feynman:1984bi}. More precisely, following \cite{Costales:2023ljr}, we introduce the clock register $\mathcal{H}_\text{C} := \text{span}_{\mathbb{C}}\qty{\ket{0},\ket{1}}^{\otimes N_\text{C}}$ spanned by $N_\text{C}$ ancillary qubits. Then, analogously to the sequence of computational states, we pick a fixed sequence of $N_\text{G} + 1$ mutually different clock states $\ket{g}_\text{C} \in \mathcal{H}_\text{C}$. Standard choices of sequences include the particle-hopping $\ket{0}^{\otimes g} \ket{1} \ket{0}^{\otimes N_\text{G} - g}$ and the domain-wall $\ket{1}^{\otimes g + 1} \ket{0}^{\otimes N_\text{G} - g}$ states, both of which require $N_\text{C} = N_\text{G} + 1$ ancillary qubits \cite{Caha:2018yhv}. In this letter we select the binary encoded states $\ket{g}_\text{D} := \ket{(g)_{N_\text{C} - 1}}\cdots\ket{(g)_0}$, where $(g)_d \in \qty{0,1}$ denotes the value of the $d$-th bit in the binary representation of $g$, therefore requiring $N_\text{C} = \lceil\log_2(N_\text{G} + 1)\rceil$ ancillary qubits \cite{Barison:2022drt}.

Having defined the clock states, the circuit-to-Hamiltonian map is then given by the initial state $\ket{\psi(0)} := \ket{0} \otimes \ket{\phi}$ and the time-independent Feynman clock Hamiltonian \cite{Feynman:1984bi, Costales:2023ljr, Caha:2018yhv}
\begin{equation}\label{eq:feynman_clock}
    H := \sum_{g=1}^{N_\text{G}} \qty(\ketbra{g}{g-1} \otimes U_g + \ketbra{g-1}{g} \otimes U^\dag_g)
\end{equation}
where, to lighten the notation, from now on we set $\hbar = c = 1$ and we omit the clock and data C/D subscripts if they can be inferred by the left/right positions of the states and operators in the tensor products. By applying (powers of) $H$ to $\ket{\psi(0)}$ we understand the purpose of the clock states, that is to track in the clock register the progress of the quantum circuit computation in the data register \cite{Costales:2023ljr}: $U_g$ gates are sequentially applied to $\ket{\phi}_\text{D}$ (due to $\ketbra{g}{g-1}_\text{C}$) or removed from $\ket{\phi}_\text{D}$ (due to $\ketbra{g-1}{g}_\text{C}$). Going to the matrix exponentiation of $H$, it can be shown that the full dynamics of the system's quantum state through time $t$ is given by \cite{Costales:2023ljr}
\begin{equation}\label{eq:feynman_solution}
    \ket{\psi(t)} := e^{-itH}\ket{\psi(0)} = \sum_{g=0}^{N_\text{G}} \alpha_g(t) \ket{g} \otimes \ket{g},
\end{equation}
where
\begin{equation}\label{eq:alphas}
\begin{split}
    \alpha_g(t) = \frac{2}{N_\text{G} + 2} \sum_{g'=0}^{N_\text{G}}&\;\; \sin(\frac{\pi (g+1)(g'+1)}{N_\text{G} + 2})\\
    &\cdot \sin(\frac{\pi (g'+1)}{N_\text{G} + 2}) e^{-2it \cos(\frac{\pi(g'+1)}{N_\text{G} + 2})}.
\end{split}
\end{equation}

To perform Z-measurements on the data register as in the circuit model, \cite{Costales:2023ljr} proposes to evolve $\ket{\psi(t)} \in \mathcal{H}_\text{C} \otimes \mathcal{H}_\text{D}$ up to the shortest time $\tau$ that maximizes the probability of collapsing onto $\ket{N_\text{G}}_\text{C}$ when the clock register is Z-measured. The evolution up to this maximal probability time $\tau := \min(\text{argmax}_{t \geq 0}(\abs{\alpha_{N_\text{G}}(t)}^2)) \approx N_\text{G}/2$ and the subsequent collapse of the clock register state are repeated until $\ket{N_\text{G}}_\text{C}$ is obtained, after which the corresponding Z-measurements can be performed on the data register state $\ket{N_\text{G}}_\text{D}$ \cite{Costales:2023ljr}.

\subsection{Alternative viewpoint on Z-measurements}\label{subsec:alternativeZ}
In this letter, we propose the different viewpoint of interpreting the above Z-measurements as time-dependent expectation values of single Pauli-Z operators acting on the data register, in conjunction with projecting the clock register on the final computational state $\ket{N_\text{G}}_\text{C}$.

Let us first introduce notations to formally state the above. We uniquely label all qubits spanning $\mathcal{H}_\text{C} \otimes \mathcal{H}_\text{D}$ with an integer $i \in S := \qty{1, \dots, N_\text{C} + N_\text{Q}}$, where $S$ represents the whole chain of qubits. We also define the sets $S_\text{C} := \qty{1, \dots, N_\text{C}}$ and $S_\text{D} := \qty{N_\text{C} + 1, \dots, N_\text{C} + N_\text{Q}}$, so that $S = S_\text{C} \cup S_\text{D}$. Then, any element $A \in \mathcal{P}(S)$ of the power set of $S$ represents a subsystem. We act on the qubits selected by $A$ with Pauli strings
\begin{equation}\label{eq:pauli_string}
    \sigma^a_{A} := \prod_{i \in A} \sigma^{a_i}_i,
\end{equation}
where $a = a_A: A \to \qty{1,2,3}$ is a function that specifies the direction $a_i = a_A(i)$ of the Pauli operator $\sigma^{a_i}_i$ acting on qubit $i$. Hence, denoting $\mathcal{R}(S)$ the set of all functions from any element of $\mathcal{P}(S)$ to $\qty{1,2,3}$, any tuple $(A,a) \in \mathcal{P}(S) \times \mathcal{R}(S) =: \mathbb{P}(S)$ uniquely defines a Pauli string as in \eqref{eq:pauli_string}.

The viewpoint we take in this letter is to consider the time evolution of the following $N_\text{Q}$ quantities
\begin{equation}\label{eq:quantities_of_interest}
    \text{proj}_{N_\text{G}} \otimes \sigma^3_i = \frac{1}{2^{N_\text{C}}}\sum_{(C,c)\in \mathbb{P}(S_\text{C})} \mel{N_\text{G}}{\sigma_C^c}{N_\text{G}}_\text{C} \qty(\sigma_C^c \otimes \sigma^3_i),
\end{equation}
where $i \in S_\text{D}$ and the projector $\text{proj}_g := \ketbra{g}{g}_\text{C}$ is expanded using the Pauli decomposition formula \cite{Nielsen:2012yss}
\begin{equation}\label{eq:pauli_decomposition}
    \ketbra{g'}{g}_\text{C} = \frac{1}{2^{N_\text{C}}}\sum_{(C,c)\in \mathbb{P}(S_\text{C})} \mel{g}{\sigma_C^c}{g'}_\text{C} \sigma_C^c.
\end{equation}
Quantities \eqref{eq:quantities_of_interest} allow us to extract the Z-measurements we are interested in because, using \eqref{eq:feynman_solution} and knowing \eqref{eq:alphas}, it is at any time
\begin{equation}\label{eq:connection}
    \expval{\text{proj}_{N_\text{G}} \otimes \sigma^3_i} = \abs{\alpha_{N_\text{G}}(t)}^2 \mel{N_\text{G}}{\sigma^3_i}{N_\text{G}}_\text{D},
\end{equation}
where, from now on and for notational convenience, we omit to indicate all time dependencies in the expectation values. The $\mel{N_\text{G}}{\sigma^3_i}{N_\text{G}}_\text{D}$ expectation value represents a Z-measurement on the $i$-th data register qubit. This is because the end-of-computation state $\ket{N_\text{G}}_\text{D}$ is either (proportional to) a computational basis element of $\mathcal{H}_\text{D}$ or it is in a superposition of such basis states. In the former case, any $i$-th qubit Z-measurement yields the same $\pm 1$ fixed value, hence expectation values are equal to that value. In the latter case, any $i$-th qubit Z-measurement yields a different random $\pm 1$ value, and as a result corresponding expectation values have to be computed.

Importantly, the number of Pauli strings $\abs{\mathbb{P}(S_\text{C})}$ required in the decomposition \eqref{eq:pauli_decomposition} is polynomial in $N_\text{C}$. This is a consequence of our choice of clock states, namely $N_\text{C} = \lceil\log_2(N_\text{G} + 1)\rceil$ implies $N_\text{G} + 1 \leq 2^{N_\text{C}} \leq 2(N_\text{G} + 1)$. As a result, \eqref{eq:pauli_decomposition} sums over $4^{N_\text{C}} = (2^{N_\text{C}})^2$ Pauli strings, an amount of terms which scales as $\abs{\mathbb{P}(S_\text{C})} \sim \text{poly}(N_\text{G})$.

We collect in the set $\mathcal{Q}_0$ all Pauli strings $\sigma_C^c \otimes \sigma^3_i$ appearing on the right-hand side of all $N_\text{Q}$ quantities \eqref{eq:quantities_of_interest}, after which we uniquely label all elements of $\mathcal{Q}_0$ with an integer $q \in \qty{1, \dots, \abs{\mathcal{Q}_0}}$. The expectation values of all quantities \eqref{eq:quantities_of_interest} are therefore completely determined by the expectation values of all the elements of $\mathcal{Q}_0$, which we call the quantities of interest, and clearly $\abs{\mathcal{Q}_0} = N_\text{Q}\abs{\mathbb{P}(S_\text{C})} \sim N_\text{Q} \text{poly}(N_\text{G})$.

Denote with $\bar{x}_{qs}$ the expectation value over $N_\text{S}$ shots of the $q$-th quantity of interest at time $t_s := s \Delta t$, where $\Delta t := T/N_\text{T}$ is the constant time step between $N_\text{T} + 1$ measurements, labeled with $s \in \qty{0, \dots, N_\text{T}}$, along the time evolution window $[0, T]$. Choices of $T$ include the period of $\alpha_{N_\text{G}}(t)$ (if it is periodic) or the maximal probability time $\tau$. The discretization of the time window allows us to get rid of the time dependency in \eqref{eq:connection} by summing over all time points
\begin{equation}\label{eq:connection_sum}
    \mel{N_\text{G}}{\sigma^3_i}{N_\text{G}}_\text{D} = \frac{\sum_{s=0}^{N_\text{T}}\expval{[\text{proj}_{N_\text{G}} \otimes \sigma^3_i](t_s)}}{\sum_{s=0}^{N_\text{T}}\abs{\alpha_{N_\text{G}}(t_s)}^2},
\end{equation}
so that the value of the Z-measurement is truly independent from the arbitrary choice of time $t_s$ when applying \eqref{eq:connection}. Moreover, it could happen that $\alpha_{N_\text{G}}(t_s) \approx 0$ for a specific $t_s$, leading to a greater statistical error associated to the Z-measurement if obtained from \eqref{eq:connection}.

Measurements of $\bar{x}_{qs}$ could be obtained by Trotterization using Hamiltonian \eqref{eq:feynman_clock}, however that would lead to a significant quantum noise affecting $\bar{x}_{qs}$. This is because the execution of the large number of quantum gates making up a single Trotter slice would additionally be repeated up to $N_\text{T}$ times, amplifying the quantum error propagation \cite{Avtandilyan2024}.

In this letter, we instead recover estimations of $\bar{x}_{qs}$ performing Hadamard tests \cite{Lin:2022vrd} over partial executions of the original $U$ circuit. The obvious advantage over Trotterization is in the reduction of quantum noise, as the number of quantum gates needed for a single Hadamard test is bounded by the number $N_\text{G}$ of gates in $U$ (plus the constant additional Hadamard test gates). More specifically, again using \eqref{eq:feynman_solution}, the quantities of interest appearing on the right-hand side of \eqref{eq:quantities_of_interest} are recast into
\begin{equation}\label{eq:hadamard_decomposition}
    \expval{\sigma_C^c \otimes \sigma^3_i} = \sum_{g=0}^{N_\text{G}} \sum_{g'=0}^{N_\text{G}} \alpha^*_g(t) \alpha_{g'}(t) \mel{g}{\sigma_C^c}{g'}_\text{C} \mel{g}{\sigma^3_i}{g'}_\text{D}.
\end{equation}
Since $\mel{g}{\sigma_C^c}{g'}_\text{C}$ and \eqref{eq:alphas} can be analytically computed at any time $t$, only measurements of $\mel{g}{\sigma^3_i}{g'}_\text{D}$ are required. These are obtained with two Hadamard tests (see Appendix \ref{app:hadamard} for their practical implementation), respectively providing measurements of $\Re[\mel{\Phi}{W_i}{\Phi}]$ and $\Im[\mel{\Phi}{W_i}{\Phi}]$, where
\begin{equation}\label{eq:hadamard_test_cases}
    (\ket{\Phi}, W_i) = \begin{cases}
        (\ket{g}, \;\sigma^3_i U_{g'} \cdots U_{g + 1}), &\text{if $g < g'$}\\
        (\ket{g}, \;\sigma^3_i), &\text{if $g = g'$}\\
        (\ket{g'}, U^\dag_{g' + 1} \cdots U^\dag_{g} \sigma^3_i), &\text{if $g > g'$}
    \end{cases}.
\end{equation}
Once all $2N_\text{Q} (N_\text{G} + 1)^2$ (remember that $g,g' \in \qty{0, \dots, N_\text{G}}$, $i \in \qty{1,\dots, N_\text{Q}}$ and that $\mel{g}{\sigma^3_i}{g'}_\text{D}$ is a complex number) Hadamard test results are retrieved and stored in a classical memory, we employ \eqref{eq:hadamard_decomposition} to compute the $(N_\text{T} + 1) \abs{\mathcal{Q}_0}$ values of all $\bar{x}_{qs}$ at every time $t_s$. Overall, the quantum and classical computational overheads of our proposed procedure efficiently scale as, respectively, $N_\text{Q}\text{poly}(N_\text{G})$ and $(N_\text{T} + 1)N_\text{Q}\text{poly}(N_\text{G})$.

\subsection{BBGKY Informed Sampling Mitigation}\label{subsec:bbgkism}

We now tackle the task of mitigating the quantum error arising in the execution of the Hadamard tests. This is a critical issue because the noise of all $\mel{g}{\sigma^3_i}{g'}_\text{D}$ estimations in \eqref{eq:hadamard_decomposition} propagates into the noise of all estimations of $\expval{\sigma_C^c \otimes \sigma^3_i}$ (equivalently all $\bar{x}_{qs}$), which itself propagates into the noise of all $\expval{\text{proj}_{N_\text{G}} \otimes \sigma^3_i}$ estimations in \eqref{eq:quantities_of_interest}, ultimately propagating into the noise of the $\mel{N_\text{G}}{\sigma^3_i}{N_\text{G}}_\text{D}$ Z-measurements in \eqref{eq:connection_sum}. To reduce quantum error, we employ the BBGKY-ISM scheme introduced in \cite{Saporiti:2026lbg}, of which we now give a brief overview.

The cornerstone idea behind the BBGKY-ISM approach is to sample mitigations $x_{qs}$ of $\bar{x}_{qs}$ from a physics-informed probability distribution, which encodes a selected portion of the noiseless dynamics and effectively constrains the sampling. The latter happens because we assign a higher drawing probability to those mitigations that better satisfy the selected portion of ideal dynamics, encoded in the form of a BBGKY-like hierarchy of equations of motion. The BBGKY equation associated to a Pauli string $\sigma_A^a$ is \cite{Saporiti:2026lbg}
\begin{equation}\label{eq:bbgky}
\begin{split}
    &\dv{t} \ev{\sigma_A^a} = \sum_{(B,b) \in \mathcal{B}} 2 (-1)^{(d_{AB}^{ab} - 1)/2} (d_{AB}^{ab} \text{ mod } 2) h^b_B\\
    &\qquad\cdot\ev{\prod_{i \in A \setminus B} \sigma^{a_i}_i \prod_{j \in B \setminus A} \sigma^{b_j}_j \prod_{k \in A \cap B} \qty(\delta_{a_k b_k} + \eps_{a_k b_k c} \sigma_k^c)}.
\end{split}
\end{equation}
In the above, $\delta$ and $\eps$ are respectively the 2-dimensional Kronecker delta and the 3-dimensional Levi-Civita symbol, $d_{AB}^{ab} := \sum_{k \in A \cap B} (1 - \delta_{a_k b_k})$ is the number of different directions among common qubits of $A$ and $B$, and $h^b_B$ is a constant coefficient whose origin we now explain together with $\mathcal{B}$. Using \eqref{eq:pauli_decomposition}, we decompose the terms of the Feynman clock Hamiltonian \eqref{eq:feynman_clock} as
\begin{equation}
\begin{split}
    &\ketbra{g}{g-1} \otimes U_g + \ketbra{g-1}{g} \otimes U^\dag_g=\\
    &=\frac{1}{2^{N_\text{C}}}\sum_{(C,c)\in \mathbb{P}(S_\text{C})} \sigma_C^c \otimes \underbrace{\qty(\mel{g-1}{\sigma_C^c}{g}_\text{C}U_g + \text{h.c.})}_{=:(\Tilde{\sigma}_C^c)_g}.
\end{split}
\end{equation}
Because all $U_g$ are elements of a universal quantum gate set $\mathcal{S}$, there exists a fixed maximal number of qubits $n$ (common choices of $\mathcal{S}$ lead to $n=2$) any $U_g \in \mathcal{S}$ can act on \cite{Kitaev_1997, Nielsen:2012yss}. As a consequence, the Pauli decomposition of the Hermitian $(\Tilde{\sigma}_C^c)_g$ in elements of $\mathbb{P}(S_\text{D})$ contains at most $4^n$ Pauli strings. Grouping inside the set $\mathcal{B} \subseteq \mathbb{P}(S)$ the tensor products of all the Pauli strings stemming from the two previous decompositions, we write \eqref{eq:feynman_clock} as $H = \sum_{(B,b) \in \mathcal{B}} h^b_B \sigma^b_B$, where $h^b_B$ is the (constant in time) coefficient associated to $\sigma^b_B$. Importantly, $\abs{\mathcal{B}} = 4^n N_\text{G} \abs{\mathbb{P}(S_\text{C})} \sim \text{poly}(N_\text{G})$.

As studied in \cite{Saporiti:2026lbg}, the set of all possible $4^\abs{S}$ BBGKY equations is a hierarchy in that the Pauli string correlators on the right-hand side of \eqref{eq:bbgky} are also associated to their own BBGKY equations with their own right-hand side Pauli string correlators, leading to a hierarchical structure of such connections. In this construction, we label by $\mathcal{Q}_r$ the set of all Pauli strings connected that way to all quantities of interest $\mathcal{Q}_0$ by at most $r$ connections. Fixing a radius $r$, we extend the unique labeling of Pauli strings with the integer $q$ from $\mathcal{Q}_0$ to $\mathcal{Q}_{r+1}$, namely $q \in \qty{1, \dots, \abs{\mathcal{Q}_{r+1}}}$. This is because one needs to know the expectation values of all elements of $\mathcal{Q}_{r+1}$ in order to evaluate all BBGKY equations picked by the elements of $\mathcal{Q}_r$. Then, $z := \abs{\mathcal{Q}_r}/\abs{\mathcal{Q}_{r+1}} \in [0,1]$ quantifies the self-consistency (in terms of the number of unknowns versus the number of equations) of the selected portion of the hierarchy. Finally, since $\abs{\mathcal{Q}_0} \sim N_\text{Q}\text{poly}(N_\text{G})$ and $\abs{\mathcal{B}} \sim \text{poly}(N_\text{G})$, the efficient scaling of the BBGKY-ISM classical overhead is guaranteed in that, using a result of \cite{Saporiti:2026lbg}, it is $\abs{\mathcal{Q}_r} \sim N_\text{Q}\text{poly}(N_\text{G})$.

We now explain how the random sampling of physics-informed mitigations is performed by the BBGKY-ISM scheme \cite{Saporiti:2026lbg}. First, group together all the $x_{qs}$ with $s > 0$ into a so-called configuration $\va{x} := ((x_{11},\dots,x_{1N_\text{T}}), \dots, (x_{\abs{\mathcal{Q}_{r+1}} 1},\dots,x_{\abs{\mathcal{Q}_{r+1}} N_\text{T}}))$, where all $s=0$ initial conditions are fixed to $x_{q0} = \bar{x}_{q0}$. Next, quantify the physicality of the discretized dynamics described by $\va{x}$ with an action $S: \mathbb{R}^{\abs{\mathcal{Q}_{r+1}}N_\text{T}} \to \mathbb{R}$, whose full form is given in Appendix \ref{app:action}, so that $S(\va{x})$ is minimal when $\va{x}$ approximates the noiseless dynamics. With $S(\va{x})$ at hand, define the (unnormalized) probability density function (PDF) $\omega(\lambda, \va{x}) := \exp(-\lambda S(\va{x}))$, where $\lambda \geq 0$ is an arbitrary parameter relating to the widths of the PDF probability peaks. Given a probability distribution defined by this PDF, the configurations most likely to be drawn from it are those minimizing the action, which are exactly the sought-after physically-motivated mitigations of $\bar{x}_{qs}$. Finally, we sample configurations by generating a Markov Chain of $M$ realizations $\va{x}_{m \in \qty{1,\dots,M}}$ in a Simulated Annealing procedure as in \cite{Saporiti:2026lbg}. Succinctly, we start from $\va{x}_0 = ((\bar{x}_{11},\dots,\bar{x}_{1N_\text{T}}), \dots, (\bar{x}_{\abs{\mathcal{Q}_{r+1}} 1},\dots,\bar{x}_{\abs{\mathcal{Q}_{r+1}} N_\text{T}}))$ and perform $M$ sweeps $\va{x}_m \to \va{x}_{m+1}$ by randomly altering the components $x^m_{qs}$ of $\va{x}_m$ and accepting the outcome $\va{x}_{m+1}$ with probability $\min(1, \exp{-\lambda[S(\va{x}_{m+1}) - S(\va{x}_m)]})$, while increasing $\lambda$ with a chosen $\Delta \lambda > 0$ at every sweep. An example of such a realization $\va{x}_m$ can be seen in the upcoming Fig.~\ref{fig:draw_mc}. By averaging the entries $x^m_{qs}$ of $M_\text{S}$ configurations samples, evenly spaced across the chain and picked after the occurrence of thermalization at sweep $M_\text{T}$, a numerical mitigation for all $\bar{x}_{qs}$ noisy measurements is obtained.

\section{Results of the mitigation}\label{sec:results}

\begin{figure}
\centering
\begin{quantikz}
\lstick{$\ket{0}$}&\gate{R_Y(\theta)}&\ctrl{1}&\rstick[2]{$\;\cos(\frac{\theta}{2})\ket{00} + \sin(\frac{\theta}{2})\ket{11}$}\\
\lstick{$\ket{0}$}&&\targ{}&
\end{quantikz}
\caption{The tunable Bell state preparation circuit $U = U(\theta)$, acting on $N_\text{Q} = 2$ qubits $\ket{\phi}_\text{D} = \ket{00}$ and composed of $N_\text{G} = 2$ gates, $U_1 = R_Y(\theta)$ and $U_2 = \text{CNOT}$. Both first- and second-qubit Z-measurements have value $\cos(\theta/2)^2 - \sin(\theta/2)^2 = \cos(\theta)$.}
\label{fig:circuit}
\end{figure}
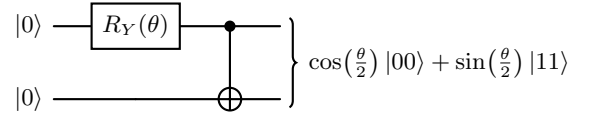

We numerically assess the effectiveness of our proposed method by applying it on the concrete example of the tunable Bell state preparation circuit, depicted in Fig.~\ref{fig:circuit}, focusing on the mitigation of its first-qubit (data register) Z-measurement. The utility of this $N_\text{Q} = N_\text{G} = 2$ circuit $U = U(\theta)$, parametrized by an angle $\theta$, resides in its use as a hardware-efficient ansatz layer for variational quantum algorithms \cite{Kandala:2017vok, 10.3389/frqst.2023.1273581}.

We execute circuit $U$ and its corresponding Hadamard tests over $N_\text{S} = 10^4$ shots in a classically simulated quantum device whose noise model mimics the current state-of-the-art level of quantum error, namely the recorded noise affecting the IBM Fez processor \footnote{The online noise model snapshot was retrieved from the IBM Quantum Platform on June 30th 2026.}. As explained in Sec.~\ref{subsec:alternativeZ}, we employ the Hadamard test results to reconstruct the time evolution of a Feynman's quantum computer execution with initial state $\ket{0} \otimes \ket{\phi}_\text{D} = \ket{0000}$. Given that \eqref{eq:alphas} applied to the $N_\text{G} = 2$ case  yields the periodic function $\alpha_{N_\text{G}}(t) = -\sin(t/\sqrt{2})^2$, of period $\sqrt{2} \pi \approx 4.5$ and whose maximal probability time $\tau$ is half of the former, we select the evolution time $T = 4.5$ and we discretize the evolution window over $N_\text{T} = 45$ time steps. Regarding the chosen BBGKY-ISM parameters, based on numerical experiments, we chose to generate a Markov Chain of $M = 2 \cdot 10^4$ configurations, starting from $\lambda = 0$ and increasing it by $\Delta \lambda = 0.5$ at every sweep, of which we select $M_\text{S} = 50$ samples after the thermalization sweep $M_\text{T} = M/2 = 10^4$. Finally, the angle $\theta \in \qty{\theta_k := (k/5)(\pi/2) : 0 \leq k \leq 4}$ and radius $r$ parameters are varied through simulations.

\begin{figure}
\centering
\includegraphics[width=\linewidth]{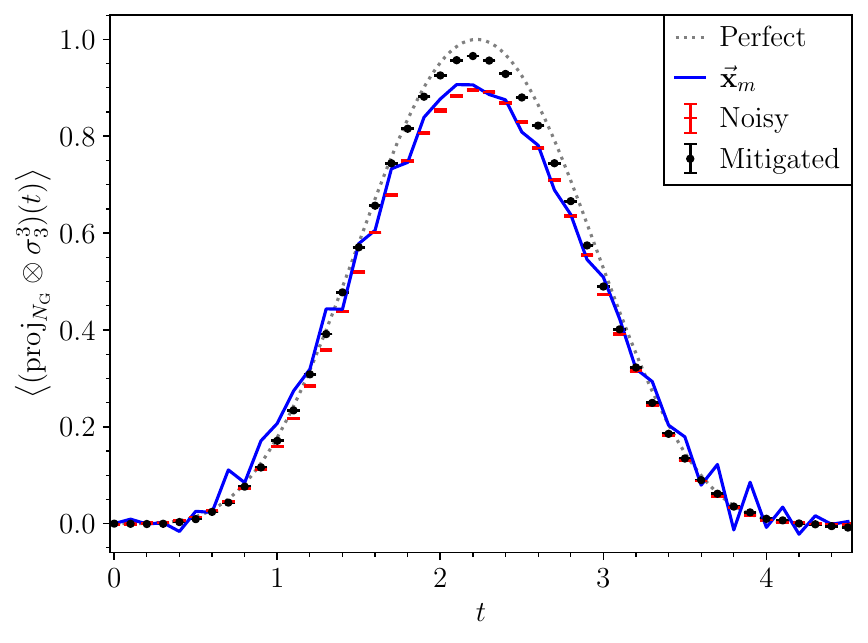}
\caption{Mitigation of the first \eqref{eq:quantities_of_interest} quantity representing the first-qubit Z-measurement of $U(\theta = 0)$ with $r = 3$. The dotted gray line is the analytical evolution computed with \eqref{eq:feynman_solution}, the red data points are obtained with the Hadamard test results as explained in Sec.~\ref{subsec:alternativeZ}, the black data points are the final results of the mitigation, and the blue line represents an example of a configuration in the Simulated Annealing procedure outlined in Sec.~\ref{subsec:bbgkism}.}
\label{fig:draw_mc}
\end{figure}

In Fig.~\ref{fig:draw_mc} we display a snapshot of the mitigation procedure performed by the BBGKY-ISM scheme, specifically on the $U(\theta = 0)$ circuit at $r = 3$. By respectively combining the noisy $\bar{x}_{qs}$ and the (averaged over $M_\text{S}$ selected indices $m$) $x^m_{qs}$ components as in \eqref{eq:quantities_of_interest}, corresponding noisy and mitigated measurements of $\expval{\text{proj}_{N_\text{G}} \otimes \sigma^3_3}$ are obtained through all time points. Both latter measurements are displayed in Fig.~\ref{fig:draw_mc} as, respectively, red and black data points. We also see from Fig.~\ref{fig:draw_mc} a clear depiction of relation \eqref{eq:connection} in that the noisy and mitigated measurements of $\expval{\text{proj}_{N_\text{G}} \otimes \sigma^3_3}$ are proportional to their respective noisy and mitigated first-qubit Z-measurements. As explained in Sec.~\ref{subsec:alternativeZ}, the latter are extracted by applying \eqref{eq:connection_sum} to their corresponding $\expval{\text{proj}_{N_\text{G}} \otimes \sigma^3_3}$ time evolution series.

\begin{figure}
\centering
\includegraphics[width=\linewidth]{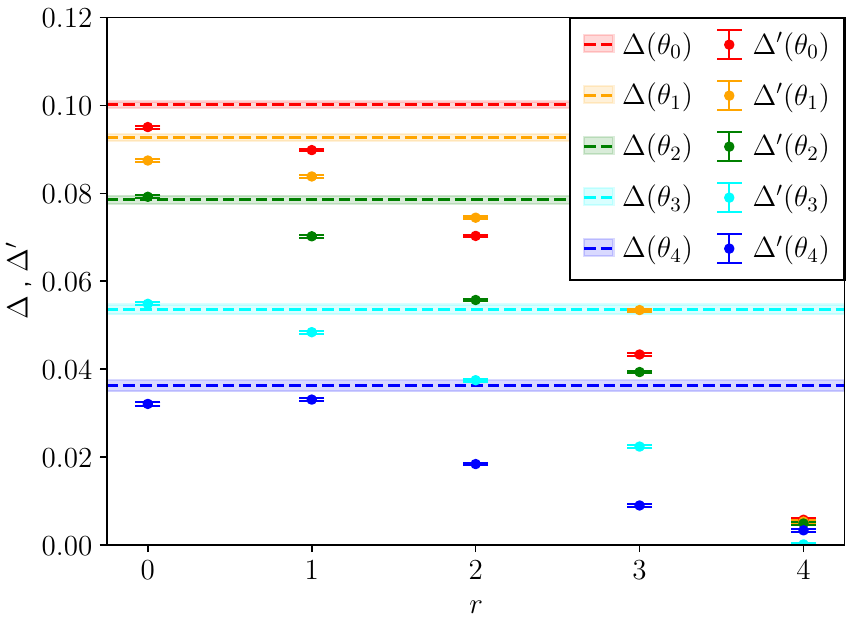}
\caption{Quantum error reduction of all first-qubit Z-measurements for different choices of $r$ and $\theta = \theta_k$. Each dashed line and data points color is associated to a corresponding $\theta_k$.}
\label{fig:draw_bell}
\end{figure}

Figure \ref{fig:draw_bell} displays the quantum error reduction capability of our proposed method as a function of the chosen radius $r$ over different chosen Bell state preparation circuits $U(\theta)$. More specifically, we define the two deltas $\Delta = \Delta(\theta)$ and $\Delta' = \Delta'(\theta)$ as the absolute differences between the theoretical first-qubit Z-measurement $\cos(\theta)$ and, respectively, the empirically measured noisy and mitigated first-qubit Z-measurements. Because they are unaffected by BBGKY-ISM and they are hence independent of $r$, the $\Delta$ are represented as lines (with corresponding error bands) rather than data points. First, we observe a systematic noise reduction $\Delta'(\theta_k) < \Delta(\theta_k)$ for all $\theta = \theta_k$ at all radii $r$, making our method effective even with the minimal $r=0$ set of selected BBGKY equations. Moreover, we observe that the error reduction capability is tunable, that is, all $\Delta'(\theta_k)$ systematically decrease as $r$ is increased, in accordance with our previous finding in \cite{Saporiti:2026lbg}. Indeed, a larger radius $r$ translates into a bigger encoded portion of the BBGKY hierarchy in $S(\va{x})$, which in turn constrains more effectively the mitigations towards the noiseless dynamics. Furthermore, we observe that $\Delta(\theta_{k+1}) < \Delta(\theta_k)$. This is a consequence of the theoretical value of the first-qubit Z-measurement, since $\theta$ approaching $\pi/2$ makes the analytical prediction $\cos(\theta)$ align with the noisy measurements, which tend to shift towards a null value as showcased in Fig.~\ref{fig:draw_mc}. Finally, we notice that at $r=4$ it is $\Delta'(\theta_k) \approx 0$ for all $\theta = \theta_k$. This is because at that radius (and above) it is $z=1$, meaning that the system of selected (coupled differential) BBGKY equations is completely self-consistent. As a result, BBGKY-ISM ignores the quantum action contribution in $S(\va{x})$ (see Appendix \ref{app:action}) and reduces to a purely classical numerical method, fully determining the exact noiseless dynamics as if they were computed from \eqref{eq:feynman_solution}.

\section{Conclusions}\label{sec:conclusions}

In this letter we extended the applicability of the BBGKY-ISM scheme, originally designed to mitigate noisy quantum simulations of quantum spin chains, to the mitigation of noisy arbitrary quantum circuits, leading to a new mitigation technique for general quantum tasks. We then numerically tested the effectiveness of our method on the tunable Bell state preparation circuit, run under simulated state-of-the-art quantum noise.

Considering Feynman's quantum computer construction, an alternative way to execute circuits as the dynamical evolution of a corresponding quantum system, we developed an approach to efficiently reconstruct the time evolution of the latter from Hadamard tests of (partial executions of) the original circuit. We showed how the associated classical and quantum overheads of this procedure scale efficiently as a polynomial of $N_\text{Q}$, $N_\text{G}$ and $N_\text{T}$. This in turn allowed us to readily apply the BBGKY-ISM scheme, of which we gave a short overview, on the reconstructed time series, and extract from their mitigations corresponding mitigated Z-measurements of the original circuit.

We empirically assessed the error reduction capability of our method on simulated executions of different tunable Bell state preparation circuit realizations. We numerically demonstrated that not only our method systematically reduces quantum noise in every circumstance, but that its error reduction effectiveness increases with the selected portion of constraining BBGKY equations picked from the hierarchy, implying that the intensity of the mitigation provided by our scheme is controllable.

Further expansions of this work include the mitigation of more sophisticated quantum circuits, such as Grover's algorithm \cite{Grover:1996rk}, Deutsch-Jozsa's algorithm \cite{Deutsch:1992idk}, Shor's algorithm \cite{Shor:1994ihq, Shor:1994jg}, or more complex and deeper-layered variational ansätze \cite{Tilly:2021jem}. As our proposed scheme is an entirely post-processing technique, it could be incorporated in conjunction with or in succession to other existing mitigation techniques.

\begin{acknowledgments}
We thank Vasily Sazonov for his feedback and review of this letter. This work has received support from the French State managed by the National Research Agency under the France 2030 program with reference ANR-22-PNCQ-0002. We acknowledge the use of IBM Quantum services for this work. The views expressed are those of the authors, and do not reflect the official policy or position of IBM or the IBM Quantum team.
\end{acknowledgments}

\section*{Data Availability}
The source codes that generated and processed the data supporting the findings of this letter are available from the authors upon reasonable request.

\appendix

\section{Hadamard test}\label{app:hadamard}
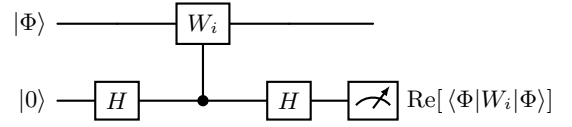
\begin{figure}[h]
\centering
\begin{quantikz}
\lstick{$\ket{\Phi}$}&&\gate{W_i}&&\\
\lstick{$\ket{0}$}&\gate{H}&\ctrl{0}\wire[u]{q}&\gate{H}&\meter{}\rstick[1]{$\Re[\mel{\Phi}{W_i}{\Phi}]$}
\end{quantikz}
\caption{Hadamard test for the real part of $\mel{\Phi}{W_i}{\Phi}$.}
\label{fig:hadamard_test}
\end{figure}

Figure \ref{fig:hadamard_test} depicts the Hadamard test circuit that, once executed with $\ket{\Phi}$ and $W_i$ as in \eqref{eq:hadamard_test_cases}, provides the real part of $\mel{\Phi}{W_i}{\Phi}$. The final Z-measurement box has to be interpreted as an expectation value over $N_\text{S}$ shots. The corresponding circuit providing the imaginary part of $\mel{\Phi}{W_i}{\Phi}$, on the other hand, is obtained by substituting the first $H$ gate acting on the ancilla qubit $\ket{0}$ with the (ordered from left to right) sequence of three gates $X$, $H$ and $S$. \cite{Lin:2022vrd}

\section{The BBGKY-ISM action}\label{app:action}
In \cite{Saporiti:2026lbg} we proposed the action
\begin{equation}
    S(\va{x}) := \qty(1 - z) S_\text{Q}(\va{x}) + z S_\text{B}(\va{x}),
\end{equation}
where $z := \abs{\mathcal{Q}_r}/\abs{\mathcal{Q}_{r+1}}$ and the BBGKY action $S_\text{B}(\va{x})$ and the quantum measurements action $S_\text{Q}(\va{x})$ are defined, respectively, as
\begin{equation}
\begin{split}
    S_\text{B}(\va{x}) &:= \frac{\abs{\mathcal{Q}_{r+1}}}{\abs{\mathcal{Q}_{r}}} \Delta t \sum_{(A,a) \in \mathcal{Q}_r} \sum_{s=0}^{N_\text{T}} E_A^a(s, \va{x})^2,\\
    S_\text{Q}(\va{x}) &:= \frac{\Delta t}{2} \sum_{q=1}^\abs{\mathcal{Q}_{r+1}} \sum_{s=0}^{N_\text{T}} \frac{(x_{qs} - \bar{x}_{qs})^2}{1-\bar{x}^2_{qs}}.
\end{split}
\end{equation}
In the above, $E^a_A(s, \va{x})$ denotes the discretization of \eqref{eq:bbgky} evaluated at time $t_s$, that is, it approximates the time derivative of \eqref{eq:bbgky} with combinations of finite difference schemes as in \cite{Saporiti:2026lbg}. Moreover, for conceptual simplicity, in this letter we assume $\abs{\bar{x}_{qs}} < 1$, as the $\abs{\bar{x}_{qs}} = 1$ cases were discussed in \cite{Saporiti:2026lbg}.


\bibliography{bio}

\end{document}